\documentclass[aps,superscriptaddress,prl,twocolumn,amsmath,amssymb]{revtex4-1}
\usepackage{graphicx}
\usepackage{hyperref}
\usepackage[usenames,dvipsnames]{xcolor}
\usepackage{eurosym}
\usepackage{dcolumn}
\usepackage{bm}
\usepackage{subfigure}
\usepackage[final]{pdfpages}
\begin{document}
%define some useful commands
\newcommand{\tr}{\operatorname{Tr}}
\newcommand{\ket}[1]{\left | #1 \right \rangle}
\newcommand{\bra}[1]{\left \langle #1 \right |}
\newcommand{\proj}[1]{\ket{#1}\!\!\bra{#1}}
%end of custom commands
\title{Quantum walks of correlated photon pairs in two-dimensional waveguide arrays}
\date{\today}
\author{Konstantinos Poulios}
\affiliation{Centre for Quantum Photonics, H. H. Wills Physics Laboratory \& Department of Electrical and Electronic Engineering, University of Bristol, Merchant Venturers Building, Woodland Road, Bristol, BS8 1UB, UK}
\author{Robert Keil}
\affiliation{Institute of Applied Physics, Friedrich-Schiller-Universit\"{a}t Jena, Max-Wien-Platz 1, D-07743 Jena, Germany}
\author{Daniel Fry}
\author{Jasmin D. A. Meinecke}
\author{Jonathan C. F. Matthews}
\affiliation{Centre for Quantum Photonics, H. H. Wills Physics Laboratory \& Department of Electrical and Electronic Engineering, University of Bristol, Merchant Venturers Building, Woodland Road, Bristol, BS8 1UB, UK}
\author{Alberto Politi}
\affiliation{Centre for Quantum Photonics, H. H. Wills Physics Laboratory \& Department of Electrical and Electronic Engineering, University of Bristol, Merchant Venturers Building, Woodland Road, Bristol, BS8 1UB, UK}
\affiliation{Now at: Center for Spintronics and Quantum Computation, University of California, Santa Barbara, California 93106, USA}
\author{Mirko Lobino}
\affiliation{Centre for Quantum Photonics, H. H. Wills Physics Laboratory \& Department of Electrical and Electronic Engineering, University of Bristol, Merchant Venturers Building, Woodland Road, Bristol, BS8 1UB, UK}
\affiliation{Now at: Centre for Quantum Dynamics and Queensland Micro and Nanotechnology Centre, Griffith University, Brisbane, QLD 4111, Australia}
\author{Markus Gr\"{a}fe}
\affiliation{Institute of Applied Physics, Friedrich-Schiller-Universit\"{a}t Jena, Max-Wien-Platz 1, D-07743 Jena, Germany}
\author{Matthias Heinrich}
\affiliation{Institute of Applied Physics, Friedrich-Schiller-Universit\"{a}t Jena, Max-Wien-Platz 1, D-07743 Jena, Germany}
\affiliation{Now at: CREOL, The College of Optics \& Photonics, University of Central Florida, 4000 Central Florida Blvd., Orlando FL 32816, USA}
\author{Stefan Nolte}
\author{Alexander Szameit}
\affiliation{Institute of Applied Physics, Friedrich-Schiller-Universit\"{a}t Jena, Max-Wien-Platz 1, D-07743 Jena, Germany}
\author{Jeremy L. O'Brien}
\affiliation{Centre for Quantum Photonics, H. H. Wills Physics Laboratory \& Department of Electrical and Electronic Engineering, University of Bristol, Merchant Venturers Building, Woodland Road, Bristol, BS8 1UB, UK}
\begin{abstract}
We demonstrate quantum walks of correlated photons in a 2D network of directly laser written waveguides coupled in a \lq swiss cross\rq \ arrangement. The correlated detection events show high-visibility quantum interference and unique composite behaviour: strong correlation and independence of the quantum walkers, between and within the planes of the cross. Violations of a classically defined inequality, for photons injected in the same plane and in orthogonal planes, reveal non-classical behaviour in a non-planar structure.
\end{abstract}
\maketitle
Quantum walks (QWs) are an illustrative example for the indeterminism inherent to quantum mechanics \cite{ah-pra-48-1687}. They are used to model a variety of processes ranging from excitation transfer across spin chains \cite{bo-prl-91-207901,ch-prl-92-187902} to energy transport in photosynthetic complexes \cite{pl-njp-10-113019}. They enable studying large scale quantum interference \cite{pe-sci-329-1500} and their simulation on a quantum computer provides a route to universal quantum computing \cite{ch-prl-102-180501}. One-dimensional (1D) networks provide a conceptually straight-forward and readily implementable way of realising QWs and highlighting their distinct differences from their classical counterparts, random walks. 

There is now a plethora of single particle QW implementations across different platforms \cite{sc-prl-103-090504, za-prl-104-100503, ka-sci-325-174, bo-pra-61-013410, ry-pra-72-062317, du-pra-77-042316, do-josab-22-499, br-prl-104-153602, sc-prl-104-050502, sch-prl-106-180403, pe-prl-100-170506}, however the dynamics of single particle QWs can be described in the context of classical wave theory \cite{kn-pra-68-020301}. QWs of multiple indistinguishable particles, on the other hand, have been shown to exhibit non-classical correlations. In general, they cannot be described by considering separately the quantum state of each particle and their features cannot be mimicked with classical light without limiting the visibility of observed quantum interference \cite{br-prl-102-253904} or the introduction of an increasing number of experiments \cite{ke-pra-81-023834,ke-pra-83-013808,sc-sci-336-55}. Furthermore, the Hilbert space that describes multi-particle QWs grows exponentially with the linear increase in particle number and network size and it has been shown that the introduction of multiple walkers increases the dimensionality of QWs \cite{pe-sci-329-1500, ro-arxiv-1205.1850}. 

An additional physical dimension in the network can, in principle, entail the degrees of freedom offered by two walkers on a 1D-network \cite{Longhi:BlochOscillationsCorrelatedParticlesSim2D,Graefe:2DSimulatorBiphotonWalk,sc-sci-336-55}. Moreover, the additional physical dimension enables network configurations (directly mapping to graph structures) that are not otherwise available, allowing for example a selective degree of connectivity for different single sites (vertices) in the structure and asymmetries in the network. Many interesting problems, such as energy transport in biological systems \cite{mo-jchemphys-129-174106,pl-njp-10-113019}, graph theory problems \cite{ha-prsa-467-212} and quantum search algorithms \cite{ch-stoc03proc-59}, require two-dimensional (2D) graphs, with a high degree of connectivity, where vertices are connected to multiple edges. 

\begin{figure}[bp]
    \centering
    \includegraphics[width = 8.6cm]{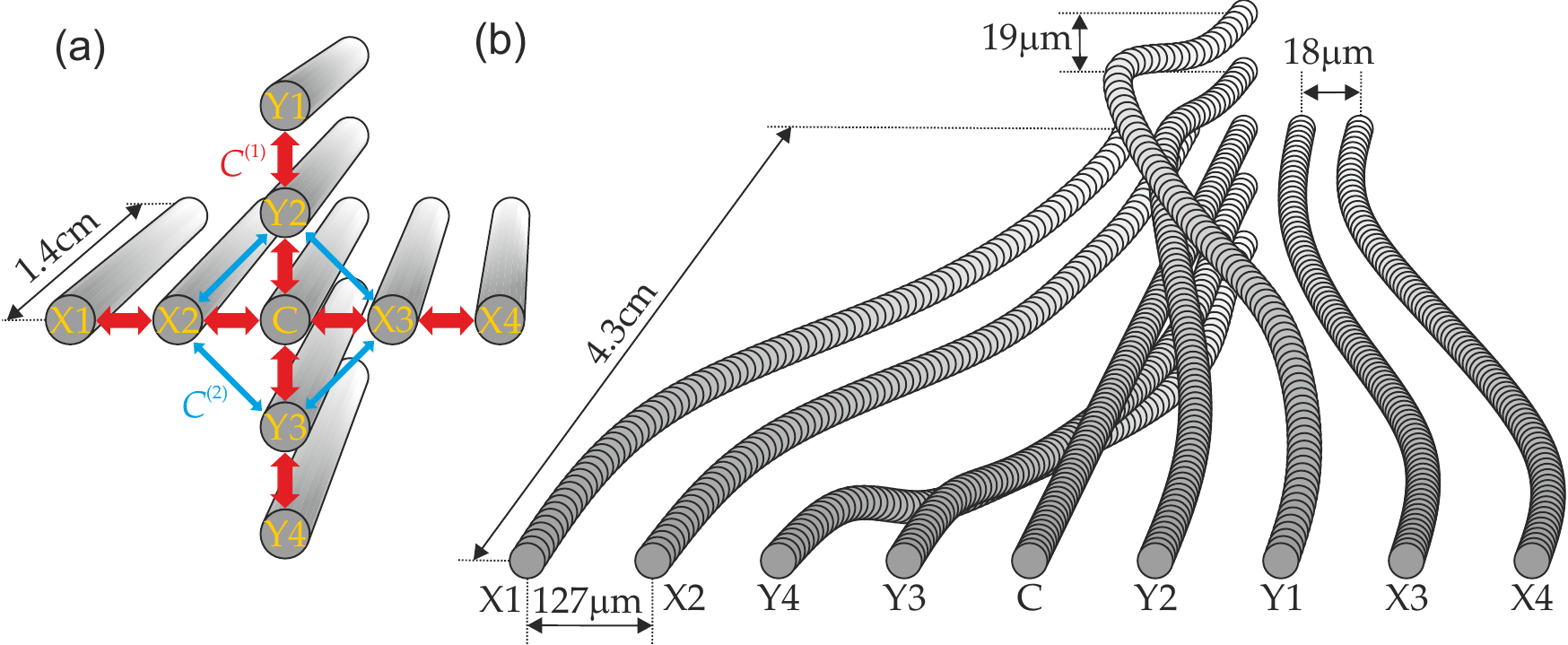}
\caption[]{\footnotesize{(a) Schematic of the 2D evanescently coupled waveguide array. The coupling constant $C^{(1)}$ is for adjacent waveguides and the second order coupling is denoted as $C^{(2)}$. (b) Schematic of the interface section of our waveguide circuit, showing the input waveguides fanning from a planar configuration (accessible with commercial fibre arrays) to the 2D, \lq swiss cross\rq \ configuration.}}
\label{fig:coupling}
\end{figure}  

A promising route towards the realisation of networks with topologies beyond 1D, nearest-neighbour coupled geometries is the direct laser inscription of waveguides in transparent substrates. This technique allows the fabrication of complex 2D networks on a single optical chip (with the third dimension representing time). It overcomes the limitations of established lithographic methods and allows implementation of intricate topologies in an integrated fashion. The quantum behaviour of single photons in simple planar arrangements of laser written waveguides, such as directional couplers, has been demonstrated with high visibility \cite{ma-oe-17-12546}. The additional dimension allowed by the direct laser write technique has also been employed to compensate for polarization dependent coupling \cite{sa-prl-108-010502} and altering the boundary conditions of 1D QWs \cite{ow-njp-13-075003}. 

In this work we implement QWs of two indistinguishable photons in a 2D waveguide lattice with sites that have a high degree of connectivity. This lattice was fabricated in fused silica with the direct laser write technique \cite{sz-jphysb-43-163001}. We observe correlations that strongly violate the limit for classical light propagating in the network, illustrating high-visibility quantum interference at the single photon level. 

The Hamiltonian for a system of N evanescently coupled waveguides is given by \cite{ke-pra-81-023834}: 
\begin{equation}
\hat{H}=\sum_{q=1}^N{\left(\beta_q a_q^{\dagger}a_q+\sum_{r=1}^N{C_{q, r}a_{r}^{\dagger}a_{q}}\right)},
\label{eq:hamiltonian}
\end{equation}
and is equivalent to the adjacency matrix of the graph representing the waveguide structure, where $a^{\left(\dagger\right)}_{q}$ is the bosonic annihilation (creation) operator for a photon in waveguide $q$, $\beta_{q}$ is the propagation constant (on site potential) of guide $q$ and $C_{q, r}$ is the coupling strength (hopping amplitude) between waveguides $q$ and $r$. For an array of uniform waveguides, $\beta_{q}$ is constant and the coupling strength between two waveguides depends solely on their distance. Note that in general each waveguide can be coupled to several other waveguides, enabling the fabrication of structures that directly map to graphs with a high degree of connectivity, with different coupling strengths for nearest and non-nearest neighbouring waveguides. From the Hamiltonian one obtains a unitary operator $\hat{U}(z)=\exp(-i \hat{H}z)$ which governs the evolution of the creation operators $a_{q'}^{\dagger}(z)=\sum_{q=1}^N{U_{q',q}a_q^{\dagger}(z=0)}$ along the longitudinal coordinate $z$, with $z$ directly translating to the time parameter $t$.

\begin{figure}[tp]
	\centering
		\includegraphics[width = 8.6cm]{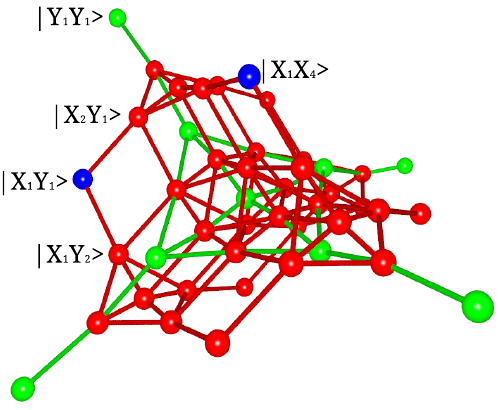}
		\caption{\footnotesize{Graph structure simulated with a two-photon input state in the \lq swiss cross\rq \ structure shown in Fig.~\ref{fig:coupling}(a). Each vertex represents a two photon state with different degrees of connectivity (up to degree 8). The on site potential for red and green vertices is $\beta$ and $2\beta$ respectively, while the hopping amplitude is $\sqrt{2}C$ for green edges in the graph and $C$ for red edges. The two blue vertices are the two different two-photon input states in the experiment (X1-X4 and X1-Y1). }}
		\label{fig:graph}
\end{figure}

For two indistinguishable input photons in waveguides $q$ and $r$, the probability of detecting one photon in output waveguide $q'$, coincident with the other photon in waveguide $r'$ is given by the correlation function \cite{ma-apb-60-s111}: 
\begin{equation}
\Gamma_{q',r'}^{(q,r)}=\frac{1}{1+\delta_{q',r'}}\left|U_{q',q}U_{r',r}+U_{q',r}U_{r',q}\right|^2.
\label{eq:corfun}
\end{equation}
In contrast, for the case of distinguishable photons no interference occurs and the probability is given by \cite{ma-apb-60-s111}: 
\begin{equation}
\Gamma_{q',r'}^{'(q,r)}=\frac{1}{1+\delta_{q',r'}}[\left|U_{q',q}U_{r',r}\right|^2+\left|U_{q',r}U_{r',q}\right|^2].
\label{eq:clascorfun}
\end{equation}
A sufficient criterion for non-classical behaviour is violation of inequality \cite{pe-sci-329-1500,br-prl-102-253904}:  
\begin{equation}
V_{q,r}=\frac{2}{3}\sqrt{\Gamma^{cl}_{q,q}\Gamma^{cl}_{r,r}}-\Gamma^{cl}_{q,r}<0
\label{eq:violeq}
\end{equation}
with $\Gamma^{cl}$ here referring to intensity correlations between classical light beams. This inequality imposes a limit to the magnitude of the on-diagonal terms in the correlation matrix (representing the presence of both photons in the same waveguide) in comparison to the associated off-diagonal elements (photons detected in different waveguides). Its violation in the quantum regime is a sign of photon-bunching. 

\begin{figure*}[htp!]
    \centering
    \includegraphics[width = 1\linewidth]{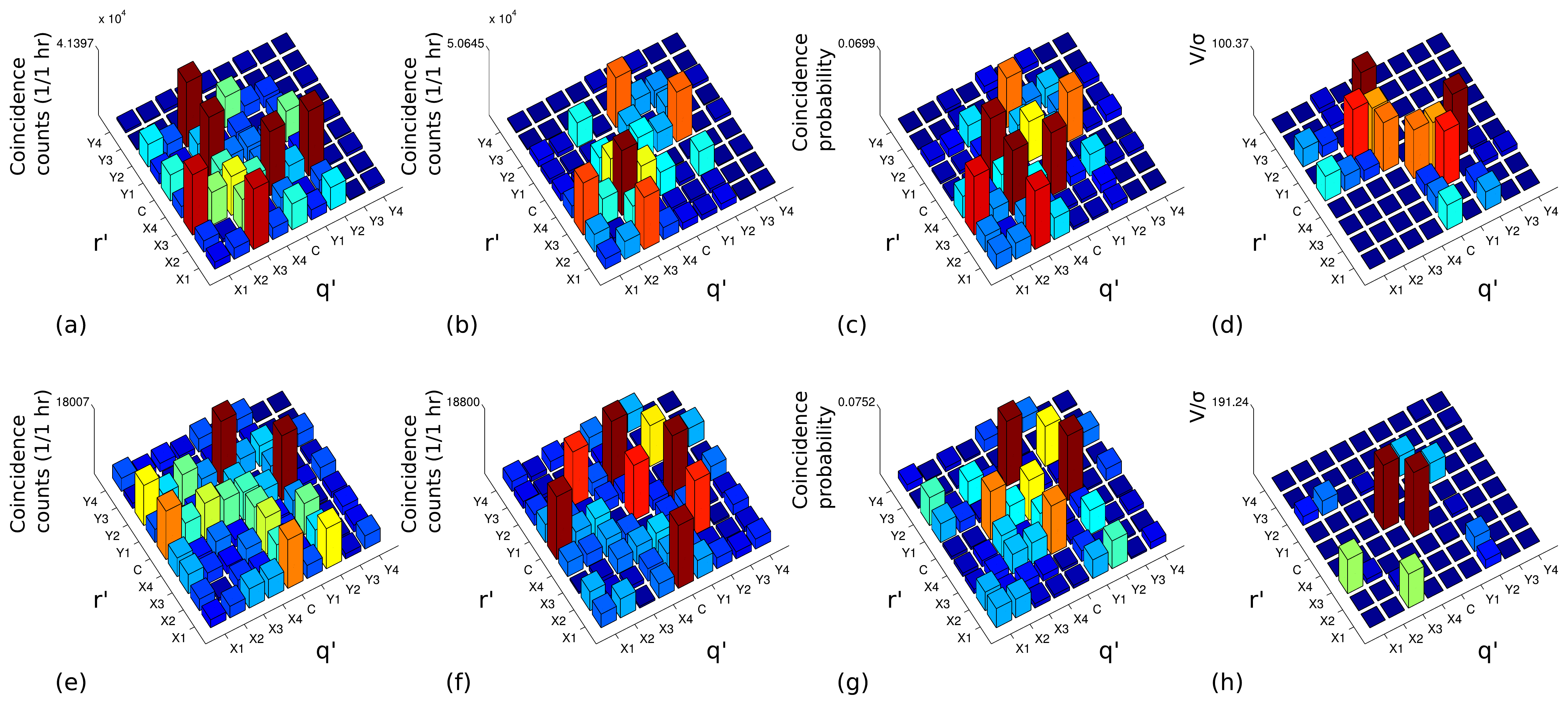}
\caption[]{\footnotesize{Correlation matrices for the two different input combinations. Top row is for inputting the two photons on the same plane in opposite corners (X1 and X4). The measured results for delays leading to distinguishable (a) and indistinguishable (b) photons are presented together with the numerical simulation (c) and the violations of inequality (\ref{eq:violeq}) (measured in standard deviations) (d), both for the case of indistinguishable photons. The bottom row shows the corresponding results for inputting the two photons in different planes in adjacent corners (X1 and Y1). For both inputs, violations of the inequality for distinguishable photons were not observed, as was expected.}}
\label{fig:corrmats}
\end{figure*}

The waveguides of the QW network in this work are labelled as in Fig.~\ref{fig:coupling}(a), where waveguides that have first order coupling in the horizontal plane are denoted with a prefix $X$, those with first order coupling in the vertical plane with $Y$. The central waveguide has first order coupling in both the horizontal and vertical plane and is labelled as C. For a single particle walk the size of the Hilbert space coincides with the size of the physical network structure. For a two-particle input, the Hilbert space grows exponentially. This two-particle configuration space can be interpreted as the Hilbert space of a single-particle QW on a more complex graph with a probability distribution equalling the original two-particle correlation function. The on site potentials and hopping amplitudes in this simulated single particle graph can be deduced from considering the Heisenberg equation of motion $\frac{d}{dz}\hat{A}(z)=[\hat{H},\hat{A}]$ for the single particle input $\hat{A}=a_q^\dagger$ and the two particle input $\hat{A}=a_q^\dagger a_r^\dagger$ as described in Ref. \cite{pe-sci-329-1500}. The single-particle graph structure corresponding to two-particles on the \lq swiss cross\rq \ structure is shown in Fig.~\ref{fig:graph}.

The waveguides were inscribed in a $10~\rm{cm}$ long glass sample, with parameters as reported in Ref. \cite{sz-jphysb-43-163001}. The waveguide separation in the actual lattice (length $1.4~\rm{cm}$) was $18~\rm{\mu m}$ in the X-plane and $19~\rm{\mu m}$ in the Y-plane, ensuring even coupling strengths of $C^{(1)}\approx 1.5~\rm{cm^{-1}}$ at the wavelength of $808~\rm{nm}$ in both directions. Our waveguide circuits were measured in a typical two-photon setup as described in Ref. \cite{pe-ncomms-2-224}. The polarization degenerate photon pairs at a wavelength of $808~\rm{nm}$ were generated by means of type-I spontaneous parametric downconversion. To allow for commercially available planar fibre arrays to be used for photon injection/collection, we designed interface sections where the individual waveguides are seamlessly rearranged between the 1D and 2D geometries. The structure (see Fig.~\ref{fig:coupling}(b)) harnesses the writing technique's 3D degrees of freedom to avoid crossings between the individual waveguides and to minimize their respective curvatures. The bending losses in the final device were measured and their uneven distribution was taken into account for the subsequent theoretical analysis.

We measured correlation matrices for two different input states, one corresponding to injection of the twin photons in waveguides located on the same plane (waveguides X1 and X4) and also in waveguides located in orthogonal planes (X1 and Y1). By varying the relative temporal delay between the two input photons, their degree of indistinguishability was tuned. Thereby, their correlation can be adjusted continuously between the extremes of Eqns. (\ref{eq:corfun}) and (\ref{eq:clascorfun}). The non-classical nature of the correlations measured can be quantified by the violations of inequality (\ref{eq:violeq}).

In the correlation matrices summarised in Fig.~\ref{fig:corrmats} one can identify four regions, two for correlated detection events between output waveguides in the same plane ($\Gamma_{X1-X4,X1-X4};\Gamma_{Y1-Y4,Y1-Y4}$) and two for events between waveguides in different planes ($\Gamma_{X1-X4,Y1-Y4};\Gamma_{Y1-Y4,X1-X4}$). From these it can be seen that the distinct features that appear for indistinguishable photons and the violations of the classical limit spread throughout the 2D network. The observed behaviour therefore cannot be attributed to independent, 1D, single photon QWs, but rather is characteristic of a single 2D QW of two correlated photons.

For a comparison of our experimental results with theory, we calculated the similarities $S=\left(\sum_{q',r'}{\sqrt{\Gamma^{exp}_{q',r'}\cdot \Gamma^{th}_{q',r'}}}\right)^2/(\sum_{q',r'}{\Gamma^{exp}_{q',r'}}\sum_{q',r'}{\Gamma^{th}_{q',r'}})$ between the experiments and simulations. For the theory matrices the propagation and coupling constants were deduced from classical light measurements via numerical optimisation and the input coupling efficiencies of single waveguides were deduced from single photon measurements via numerical optimisation. Exhibiting similarities of $93.97\%$ (injection in the same plane) and $81.7\%$ (orthogonal planes) with respect to the simulations, the experimental results are in good agreement with the theoretical predictions as shown in Figs.~\ref{fig:corrmats}(c,g). The according similarities for the correlation matrices of distinguishable photons are $98.57\%$ (for same plane input photons) and $98.45\%$ (for orthogonal planes).

\begin{figure}[htp!]
    \centering
    \includegraphics[width = 8.6cm]{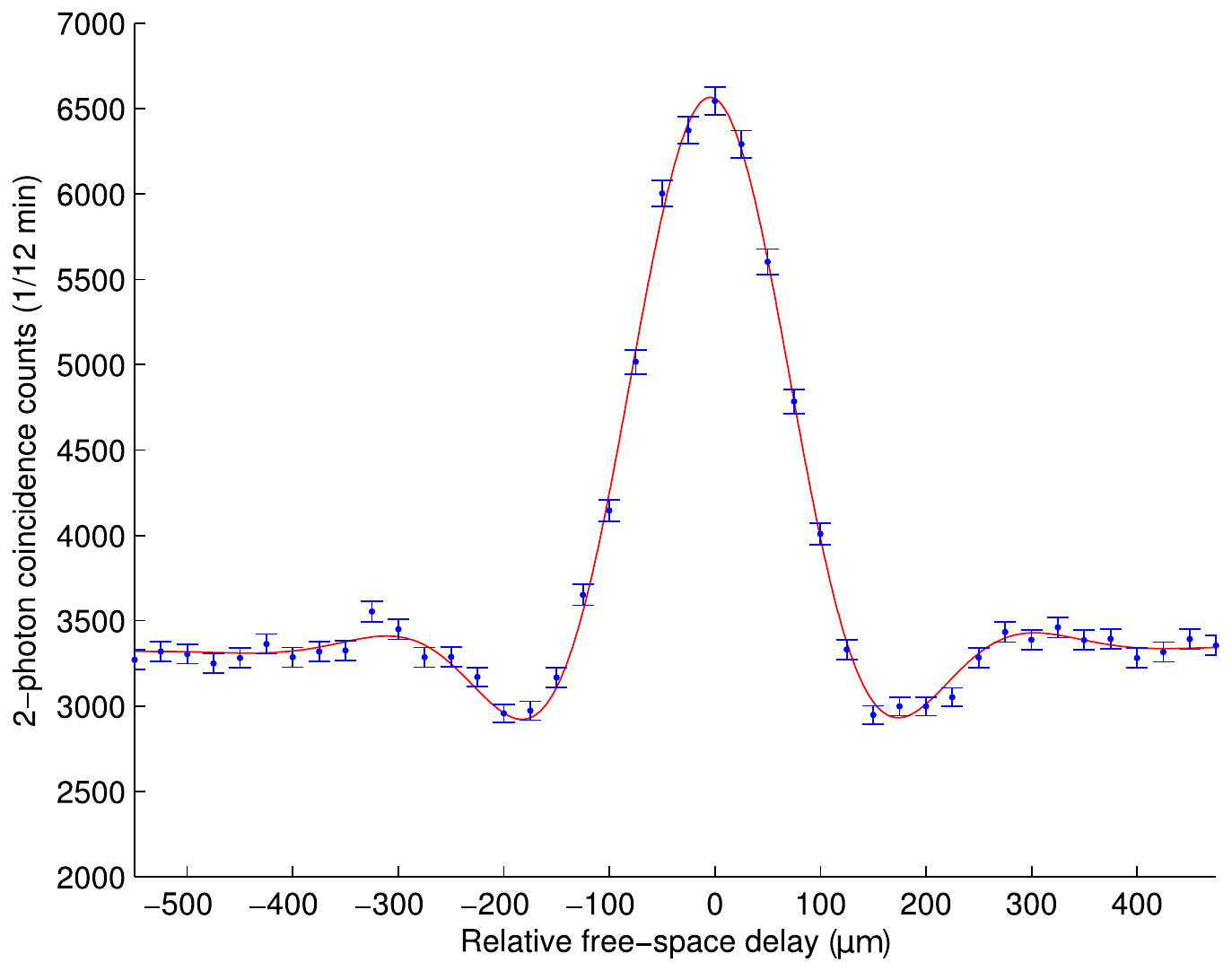}
\caption[]{\footnotesize{Quantum interference peak observed when varying the relative free-space delay of the twin photons and monitoring the presence of both photons in the same waveguide using a balanced fibre beam-splitter butt-coupled to the output waveguide. This particular peak is for both photons input on the same plane (X1, X4) and monitoring output waveguide X3, showing a visibility of $96.8\pm2.5\%$. (Error bars indicate the standard deviation of the underlying Poissonian photon counting statistics.) }}
\label{fig:hom_peak}
\end{figure}

A measure of the quality of the non-classical interference within the network can be obtained by looking at the diagonal elements of the correlation matrices, i.e., the coincidence of the two photons in the same output waveguide. These correlations are measured by using fibre-splitters after the output fibre array, thus probabilistically splitting the two photons occupying the same waveguide. The maximum visibility of the non-classical interference for these diagonal elements depends solely on the quality of the indistinguishability between the two photons inside the quantum network, irrespective of the network structure and the evolution length of the coupling region. Thus, a $100\%$ visibility peak (in this case a doubling in the coincidental count rate after the splitter) should ideally be measured when changing from distinguishable to indistinguishable photons \cite{ma-apb-60-s111}. This is evident from Eqns. (\ref{eq:corfun}) and (\ref{eq:clascorfun}) for $q'=r'$, i.e. both photons detected in the same output waveguide. Fig.~\ref{fig:hom_peak} shows one of the peaks measured with visibility $V\equiv\left(\Gamma-\Gamma^{'}\right)/\Gamma^{'}=96.8\pm2.5\%$, demonstrating the high quality of the non-classical interference in our devices. 

\begin{figure*}[htp!]
    \centering
    \includegraphics[width = 1\linewidth]{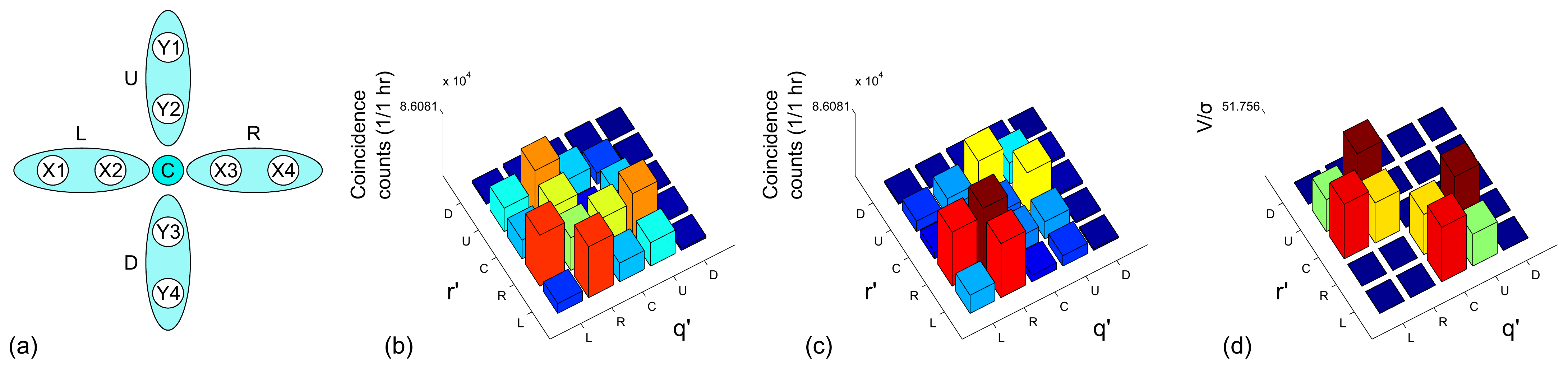}
\caption[]{\footnotesize{(a) Schematic of the 2D waveguide array when summing correlations over sites within the different branches. (b,c) Correlation matrices between the branches for inputting the two photons on the same plane in opposite corners (X1 and X4), showing the distinguishable (b) and indistinguishable (c) case. (d) Violations of inequality (\ref{eq:violeq}) expressed in standard deviations for indistinguishable photons. No violations were observed for distinguishable photons, as was expected. }}
\label{fig:summed_corrmats}
\end{figure*}

To elucidate the impact of the particular \lq swiss cross\rq \ geometry on the QW, it is instructive to investigate the correlation among the branches (left - L, right - R, up - U, down - D, as shown in Fig.~\ref{fig:summed_corrmats}(a)), as obtained by summation over their constituent waveguides' individual coincidences $\Gamma_{q',r'}^{(X1,X4)}$. Figs.~\ref{fig:summed_corrmats}(b-d) show the results of this analysis. Interestingly, no violations occur between the vertical branch (U) and the center (C), as well as between the left (L) and the right (R) branch. Moreover, the single waveguide violations matrix (see Fig.~\ref{fig:corrmats}(d)) reveals no intra-branch violations, either. In contrast, strong violations arise across the input branches (L and R) on one side and U and C on the other side. Hence, the photons tend to occupy either the input branches or the vertical branch/center. This phenomenon is reminiscent of photon bunching observed in 1D lattices \cite{pe-sci-329-1500,br-prl-102-253904}. At the same time, they are distributed almost independently, i.e., behave like distinguishable particles, within these two regions. This composite behaviour is a direct consequence of the non-planar geometry of the 2D network, confirming theoretical predictions \cite{ke-pra-81-023834}. 

In conclusion we experimentally implemented a 2D, two-photon QW by using waveguide arrays in a \lq swiss cross\rq \ geometry. We demonstrated quantum interference at the single photon level with high visibility on fully 3D integrated waveguide devices with features which cannot be observed in planar arrangements, including the combination of strong correlation and independence of the quantum walkers between and within the planes of the cross. The ability to inscribe waveguides in three dimensions enables the implementation of networks with topologies that go beyond the restriction of lithographic methods. Furthermore, QWs of multiple particles enable the simulation of graph structures which are not implementable as a physical waveguide structure. The use of entanglement to simulate different classes of particles \cite{ma-scirep-3-1539, sa-prl-108-010502} as well as precise control of the dynamics and boundary conditions of the system \cite{me-pra-88-012308} have been demonstrated on large waveguide arrays. These results and the possibility of introducing interactions between particles (e.g. measurement-induced \cite{kn-nature-409-46} or simulating 1D interactions with 2D waveguide lattices \cite{co-ncomm-4-1555}), open ways for new, interesting experiments in the quantum regime, such as universal quantum computing \cite{ch-sci-339-791}. 

\footnotesize{The authors would like to thank Lana Beck for helpful discussions. This work was supported by EPSRC, ERC, IARPA, QUANTIP, PHORBITECH, BMBF (ZIK 03Z1HN31 `Ultra Optics 2015') and TMBWK (Research group `Spacetime', Grant no. 11027-514). R.K. was supported by the Abbe School of Photonics. J.C.F.M. is supported by a Leverhulme Trust Early Career Fellowship. M.H. was supported by the German National Academy of Sciences Leopoldina (grant No. LPDS 2012-01). J.L.OB. acknowledges a Royal Society Wolfson Merit Award and a Royal Academy of Engineering Chair in Emerging Technologies. }


\begin{thebibliography}{39}%
\makeatletter
\providecommand \@ifxundefined [1]{%
 \@ifx{#1\undefined}
}%
\providecommand \@ifnum [1]{%
 \ifnum #1\expandafter \@firstoftwo
 \else \expandafter \@secondoftwo
 \fi
}%
\providecommand \@ifx [1]{%
 \ifx #1\expandafter \@firstoftwo
 \else \expandafter \@secondoftwo
 \fi
}%
\providecommand \natexlab [1]{#1}%
\providecommand \enquote  [1]{``#1''}%
\providecommand \bibnamefont  [1]{#1}%
\providecommand \bibfnamefont [1]{#1}%
\providecommand \citenamefont [1]{#1}%
\providecommand \href@noop [0]{\@secondoftwo}%
\providecommand \href [0]{\begingroup \@sanitize@url \@href}%
\providecommand \@href[1]{\@@startlink{#1}\@@href}%
\providecommand \@@href[1]{\endgroup#1\@@endlink}%
\providecommand \@sanitize@url [0]{\catcode `\\12\catcode `\$12\catcode
  `\&12\catcode `\#12\catcode `\^12\catcode `\_12\catcode `\%12\relax}%
\providecommand \@@startlink[1]{}%
\providecommand \@@endlink[0]{}%
\providecommand \url  [0]{\begingroup\@sanitize@url \@url }%
\providecommand \@url [1]{\endgroup\@href {#1}{\urlprefix }}%
\providecommand \urlprefix  [0]{URL }%
\providecommand \Eprint [0]{\href }%
\providecommand \doibase [0]{http://dx.doi.org/}%
\providecommand \selectlanguage [0]{\@gobble}%
\providecommand \bibinfo  [0]{\@secondoftwo}%
\providecommand \bibfield  [0]{\@secondoftwo}%
\providecommand \translation [1]{[#1]}%
\providecommand \BibitemOpen [0]{}%
\providecommand \bibitemStop [0]{}%
\providecommand \bibitemNoStop [0]{.\EOS\space}%
\providecommand \EOS [0]{\spacefactor3000\relax}%
\providecommand \BibitemShut  [1]{\csname bibitem#1\endcsname}%
\let\auto@bib@innerbib\@empty
%</preamble>
\bibitem [{\citenamefont {Aharonov}\ \emph {et~al.}(1993)\citenamefont
  {Aharonov}, \citenamefont {Davidovich},\ and\ \citenamefont
  {Zagury}}]{ah-pra-48-1687}%
  \BibitemOpen
  \bibfield  {author} {\bibinfo {author} {\bibfnamefont {Y.}~\bibnamefont
  {Aharonov}}, \bibinfo {author} {\bibfnamefont {L.}~\bibnamefont
  {Davidovich}}, \ and\ \bibinfo {author} {\bibfnamefont {N.}~\bibnamefont
  {Zagury}},\ }\href {\doibase 10.1103/PhysRevA.48.1687} {\bibfield  {journal} {\bibinfo  {journal} {Phys.
  Rev. A}\ }\textbf {\bibinfo {volume} {48}},\ \bibinfo {pages} {1687}
  (\bibinfo {year} {1993})}\BibitemShut {NoStop}%
\bibitem [{\citenamefont {Bose}(2003)}]{bo-prl-91-207901}%
  \BibitemOpen
  \bibfield  {author} {\bibinfo {author} {\bibfnamefont {S.}~\bibnamefont
  {Bose}},\ }\href {\doibase 10.1103/PhysRevLett.91.207901} {\bibfield  {journal} {\bibinfo  {journal} {Phys.
  Rev. Lett.}\ }\textbf {\bibinfo {volume} {91}},\ \bibinfo {pages} {207901}
  (\bibinfo {year} {2003})}\BibitemShut {NoStop}%
\bibitem [{\citenamefont {Christandl}\ \emph {et~al.}(2004)\citenamefont
  {Christandl}, \citenamefont {Datta}, \citenamefont {Ekert},\ and\
  \citenamefont {Landahl}}]{ch-prl-92-187902}%
  \BibitemOpen
  \bibfield  {author} {\bibinfo {author} {\bibfnamefont {M.}~\bibnamefont
  {Christandl}}, \bibinfo {author} {\bibfnamefont {N.}~\bibnamefont {Datta}},
  \bibinfo {author} {\bibfnamefont {A.}~\bibnamefont {Ekert}}, \ and\ \bibinfo
  {author} {\bibfnamefont {A.~J.}\ \bibnamefont {Landahl}},\ }\href {\doibase 10.1103/PhysRevLett.92.187902} 
  {\bibfield  {journal} {\bibinfo  {journal}
  {Phys. Rev. Lett.}\ }\textbf {\bibinfo {volume} {92}},\ \bibinfo {pages}
  {187902} (\bibinfo {year} {2004})}\BibitemShut {NoStop}%
\bibitem [{\citenamefont {Plenio}\ and\ \citenamefont
  {Huelga}(2008)}]{pl-njp-10-113019}%
  \BibitemOpen
  \bibfield  {author} {\bibinfo {author} {\bibfnamefont {M.~B.}\ \bibnamefont
  {Plenio}}\ and\ \bibinfo {author} {\bibfnamefont {S.~F.}\ \bibnamefont
  {Huelga}},\ }\href {\doibase 10.1088/1367-2630/10/11/113019} {\bibfield  {journal} {\bibinfo  {journal} {New
  J. Phys.}\ }\textbf {\bibinfo {volume} {10}},\ \bibinfo {pages}
  {113019} (\bibinfo {year} {2008})}\BibitemShut {NoStop}%
\bibitem [{\citenamefont {Peruzzo}\ \emph {et~al.}(2010)\citenamefont
  {Peruzzo}, \citenamefont {Lobino}, \citenamefont {Matthews}, \citenamefont
  {Matsuda}, \citenamefont {Politi}, \citenamefont {Poulios}, \citenamefont
  {Zhou}, \citenamefont {Lahini}, \citenamefont {Ismail}, \citenamefont
  {W\"{o}rhoff}, \citenamefont {Bromberg}, \citenamefont {Silberberg},
  \citenamefont {Thompson},\ and\ \citenamefont {O'Brien}}]{pe-sci-329-1500}%
  \BibitemOpen
  \bibfield  {author} {\bibinfo {author} {\bibfnamefont {A.}~\bibnamefont
  {Peruzzo}}, \bibinfo {author} {\bibfnamefont {M.}~\bibnamefont {Lobino}},
  \bibinfo {author} {\bibfnamefont {J.~C.~F.}\ \bibnamefont {Matthews}},
  \bibinfo {author} {\bibfnamefont {N.}~\bibnamefont {Matsuda}}, \bibinfo
  {author} {\bibfnamefont {A.}~\bibnamefont {Politi}}, \bibinfo {author}
  {\bibfnamefont {K.}~\bibnamefont {Poulios}}, \bibinfo {author} {\bibfnamefont
  {X.-Q.}\ \bibnamefont {Zhou}}, \bibinfo {author} {\bibfnamefont
  {Y.}~\bibnamefont {Lahini}}, \bibinfo {author} {\bibfnamefont
  {N.}~\bibnamefont {Ismail}}, \bibinfo {author} {\bibfnamefont
  {K.}~\bibnamefont {W\"{o}rhoff}}, \bibinfo {author} {\bibfnamefont
  {Y.}~\bibnamefont {Bromberg}}, \bibinfo {author} {\bibfnamefont
  {Y.}~\bibnamefont {Silberberg}}, \bibinfo {author} {\bibfnamefont {M.~G.}\
  \bibnamefont {Thompson}}, \ and\ \bibinfo {author} {\bibfnamefont {J.~L.}\
  \bibnamefont {O'Brien}},\ }\href {\doibase 10.1126/science.1193515} {\bibfield  {journal} {\bibinfo
  {journal} {Science}\ }\textbf {\bibinfo {volume} {329}},\ \bibinfo {pages}
  {1500} (\bibinfo {year} {2010})}\BibitemShut {NoStop}%
\bibitem [{\citenamefont {Childs}(2008)}]{ch-prl-102-180501}%
  \BibitemOpen
  \bibfield  {author} {\bibinfo {author} {\bibfnamefont {A.~M.}\ \bibnamefont
  {Childs}},\ }\href {\doibase 10.1103/PhysRevLett.102.180501} {\bibfield  {journal} {\bibinfo  {journal} {Phys.
  Rev. Lett.}\ }\textbf {\bibinfo {volume} {102}},\ \bibinfo {pages} {180501}
  (\bibinfo {year} {2009})}\BibitemShut {NoStop}%
\bibitem [{\citenamefont {Schmitz}\ \emph {et~al.}(2009)\citenamefont
  {Schmitz}, \citenamefont {Matjeschk}, \citenamefont {Schneider},
  \citenamefont {Glueckert}, \citenamefont {Enderlein}, \citenamefont {Huber},\
  and\ \citenamefont {Schaetz}}]{sc-prl-103-090504}%
  \BibitemOpen
  \bibfield  {author} {\bibinfo {author} {\bibfnamefont {H.}~\bibnamefont
  {Schmitz}}, \bibinfo {author} {\bibfnamefont {R.}~\bibnamefont {Matjeschk}},
  \bibinfo {author} {\bibfnamefont {C.}~\bibnamefont {Schneider}}, \bibinfo
  {author} {\bibfnamefont {J.}~\bibnamefont {Glueckert}}, \bibinfo {author}
  {\bibfnamefont {M.}~\bibnamefont {Enderlein}}, \bibinfo {author}
  {\bibfnamefont {T.}~\bibnamefont {Huber}}, \ and\ \bibinfo {author}
  {\bibfnamefont {T.}~\bibnamefont {Schaetz}},\ }\href {\doibase 10.1103/PhysRevLett.103.090504} {\bibfield
  {journal} {\bibinfo  {journal} {Phys. Rev. Lett.}\ }\textbf {\bibinfo
  {volume} {103}},\ \bibinfo {pages} {090504} (\bibinfo {year}
  {2009})}\BibitemShut {NoStop}%
\bibitem [{\citenamefont {Z{\"a}hringer}\ \emph {et~al.}(2010)\citenamefont
  {Z{\"a}hringer}, \citenamefont {Kirchmair}, \citenamefont {Gerritsma},
  \citenamefont {Solano}, \citenamefont {Blatt},\ and\ \citenamefont
  {Roos}}]{za-prl-104-100503}%
  \BibitemOpen
  \bibfield  {author} {\bibinfo {author} {\bibfnamefont {F.}~\bibnamefont
  {Z{\"a}hringer}}, \bibinfo {author} {\bibfnamefont {G.}~\bibnamefont
  {Kirchmair}}, \bibinfo {author} {\bibfnamefont {R.}~\bibnamefont
  {Gerritsma}}, \bibinfo {author} {\bibfnamefont {E.}~\bibnamefont {Solano}},
  \bibinfo {author} {\bibfnamefont {R.}~\bibnamefont {Blatt}}, \ and\ \bibinfo
  {author} {\bibfnamefont {C.~F.}\ \bibnamefont {Roos}},\ }\href {\doibase 10.1103/PhysRevLett.104.100503}
  {\bibfield  {journal} {\bibinfo  {journal} {Phys. Rev. Lett.}\ }\textbf
  {\bibinfo {volume} {104}},\ \bibinfo {pages} {100503} (\bibinfo {year}
  {2010})}\BibitemShut {NoStop}%
\bibitem [{\citenamefont {Karski}\ \emph {et~al.}(2009)\citenamefont {Karski},
  \citenamefont {F\"{o}rster}, \citenamefont {Choi}, \citenamefont {Steffen},
  \citenamefont {Alt}, \citenamefont {Meschede},\ and\ \citenamefont
  {Widera}}]{ka-sci-325-174}%
  \BibitemOpen
  \bibfield  {author} {\bibinfo {author} {\bibfnamefont {M.}~\bibnamefont
  {Karski}}, \bibinfo {author} {\bibfnamefont {L.}~\bibnamefont {F\"{o}rster}},
  \bibinfo {author} {\bibfnamefont {J.-M.}\ \bibnamefont {Choi}}, \bibinfo
  {author} {\bibfnamefont {A.}~\bibnamefont {Steffen}}, \bibinfo {author}
  {\bibfnamefont {W.}~\bibnamefont {Alt}}, \bibinfo {author} {\bibfnamefont
  {D.}~\bibnamefont {Meschede}}, \ and\ \bibinfo {author} {\bibfnamefont
  {A.}~\bibnamefont {Widera}},\ }\href {\doibase 10.1126/science.1174436} {\bibfield  {journal} {\bibinfo
  {journal} {Science}\ }\textbf {\bibinfo {volume} {325}},\ \bibinfo {pages}
  {174} (\bibinfo {year} {2009})}\BibitemShut {NoStop}%
\bibitem [{\citenamefont {Bouwmeester}\ \emph {et~al.}(1999)\citenamefont
  {Bouwmeester}, \citenamefont {Marzoli}, \citenamefont {Karman}, \citenamefont
  {Schleich},\ and\ \citenamefont {Woerdman}}]{bo-pra-61-013410}%
  \BibitemOpen
  \bibfield  {author} {\bibinfo {author} {\bibfnamefont {D.}~\bibnamefont
  {Bouwmeester}}, \bibinfo {author} {\bibfnamefont {I.}~\bibnamefont
  {Marzoli}}, \bibinfo {author} {\bibfnamefont {G.~P.}\ \bibnamefont {Karman}},
  \bibinfo {author} {\bibfnamefont {W.}~\bibnamefont {Schleich}}, \ and\
  \bibinfo {author} {\bibfnamefont {J.~P.}\ \bibnamefont {Woerdman}},\
  }\href {\doibase 10.1103/PhysRevA.61.013410} {\bibfield  {journal} {\bibinfo  {journal} {Phys. Rev. A}\
  }\textbf {\bibinfo {volume} {61}},\ \bibinfo {pages} {013410} (\bibinfo
  {year} {1999})}\BibitemShut {NoStop}%
\bibitem [{\citenamefont {Ryan}\ \emph {et~al.}(2005)\citenamefont {Ryan},
  \citenamefont {Laforest}, \citenamefont {Boileau},\ and\ \citenamefont
  {Laflamme}}]{ry-pra-72-062317}%
  \BibitemOpen
  \bibfield  {author} {\bibinfo {author} {\bibfnamefont {C.~A.}\ \bibnamefont
  {Ryan}}, \bibinfo {author} {\bibfnamefont {M.}~\bibnamefont {Laforest}},
  \bibinfo {author} {\bibfnamefont {J.~C.}\ \bibnamefont {Boileau}}, \ and\
  \bibinfo {author} {\bibfnamefont {R.}~\bibnamefont {Laflamme}},\ }\href {\doibase 10.1103/PhysRevA.72.062317} 
  {\bibfield  {journal} {\bibinfo  {journal} {Phys. Rev. A}\ }\textbf
  {\bibinfo {volume} {72}},\ \bibinfo {pages} {062317} (\bibinfo {year}
  {2005})}\BibitemShut {NoStop}%
\bibitem [{\citenamefont {Du}\ \emph {et~al.}(2003)\citenamefont {Du},
  \citenamefont {Li}, \citenamefont {Xu}, \citenamefont {Shi}, \citenamefont
  {Wu}, \citenamefont {Zhou},\ and\ \citenamefont {Han}}]{du-pra-77-042316}%
  \BibitemOpen
  \bibfield  {author} {\bibinfo {author} {\bibfnamefont {J.}~\bibnamefont
  {Du}}, \bibinfo {author} {\bibfnamefont {H.}~\bibnamefont {Li}}, \bibinfo
  {author} {\bibfnamefont {X.}~\bibnamefont {Xu}}, \bibinfo {author}
  {\bibfnamefont {M.}~\bibnamefont {Shi}}, \bibinfo {author} {\bibfnamefont
  {J.}~\bibnamefont {Wu}}, \bibinfo {author} {\bibfnamefont {X.}~\bibnamefont
  {Zhou}}, \ and\ \bibinfo {author} {\bibfnamefont {R.}~\bibnamefont {Han}},\
  }\href {\doibase 10.1103/PhysRevA.67.042316} {\bibfield  {journal} {\bibinfo  {journal} {Phys. Rev. A}\
  }\textbf {\bibinfo {volume} {67}},\ \bibinfo {pages} {042316} (\bibinfo
  {year} {2003})}\BibitemShut {NoStop}%
\bibitem [{\citenamefont {Do}\ \emph {et~al.}(2005)\citenamefont {Do},
  \citenamefont {Stohler}, \citenamefont {Balasubramanian}, \citenamefont
  {Elliott}, \citenamefont {Eash}, \citenamefont {Fischbach}, \citenamefont
  {Fischbach}, \citenamefont {Mills},\ and\ \citenamefont
  {Zwickl}}]{do-josab-22-499}%
  \BibitemOpen
  \bibfield  {author} {\bibinfo {author} {\bibfnamefont {B.}~\bibnamefont
  {Do}}, \bibinfo {author} {\bibfnamefont {M.~L.}\ \bibnamefont {Stohler}},
  \bibinfo {author} {\bibfnamefont {S.}~\bibnamefont {Balasubramanian}},
  \bibinfo {author} {\bibfnamefont {D.~S.}\ \bibnamefont {Elliott}}, \bibinfo
  {author} {\bibfnamefont {C.}~\bibnamefont {Eash}}, \bibinfo {author}
  {\bibfnamefont {E.}~\bibnamefont {Fischbach}}, \bibinfo {author}
  {\bibfnamefont {M.~A.}\ \bibnamefont {Fischbach}}, \bibinfo {author}
  {\bibfnamefont {A.}~\bibnamefont {Mills}}, \ and\ \bibinfo {author}
  {\bibfnamefont {B.}~\bibnamefont {Zwickl}},\ }\href {\doibase 10.1364/JOSAB.22.000499} 
  {\bibfield  {journal} {\bibinfo  {journal} {J. Opt. Soc. Am. B}\ }\textbf {\bibinfo {volume} {22}},\ \bibinfo {pages} {499}
  (\bibinfo {year} {2005})}\BibitemShut {NoStop}%
\bibitem [{\citenamefont {Broome}\ \emph {et~al.}(2010)\citenamefont {Broome},
  \citenamefont {Fedrizzi}, \citenamefont {Lanyon}, \citenamefont {Kassal},
  \citenamefont {Aspuru-Guzik},\ and\ \citenamefont
  {White}}]{br-prl-104-153602}%
  \BibitemOpen
  \bibfield  {author} {\bibinfo {author} {\bibfnamefont {M.~A.}\ \bibnamefont
  {Broome}}, \bibinfo {author} {\bibfnamefont {A.}~\bibnamefont {Fedrizzi}},
  \bibinfo {author} {\bibfnamefont {B.~P.}\ \bibnamefont {Lanyon}}, \bibinfo
  {author} {\bibfnamefont {I.}~\bibnamefont {Kassal}}, \bibinfo {author}
  {\bibfnamefont {A.}~\bibnamefont {Aspuru-Guzik}}, \ and\ \bibinfo {author}
  {\bibfnamefont {A.~G.}\ \bibnamefont {White}},\ }\href {\doibase 10.1103/PhysRevLett.104.153602} 
  {\bibfield {journal} {\bibinfo  {journal} {Phys. Rev. Lett.}\ }\textbf {\bibinfo
  {volume} {104}},\ \bibinfo {pages} {153602} (\bibinfo {year}
  {2010})}\BibitemShut {NoStop}%
\bibitem [{\citenamefont {Schreiber}\ \emph {et~al.}(2010)\citenamefont
  {Schreiber}, \citenamefont {Cassemiro}, \citenamefont {Poto{\v c}ek},
  \citenamefont {G{\'a}bris}, \citenamefont {Mosley}, \citenamefont
  {Andersson}, \citenamefont {Jex},\ and\ \citenamefont
  {Silberhorn}}]{sc-prl-104-050502}%
  \BibitemOpen
  \bibfield  {author} {\bibinfo {author} {\bibfnamefont {A.}~\bibnamefont
  {Schreiber}}, \bibinfo {author} {\bibfnamefont {K.~N.}\ \bibnamefont
  {Cassemiro}}, \bibinfo {author} {\bibfnamefont {V.}~\bibnamefont {Poto{\v
  c}ek}}, \bibinfo {author} {\bibfnamefont {A.}~\bibnamefont {G{\'a}bris}},
  \bibinfo {author} {\bibfnamefont {P.~J.}\ \bibnamefont {Mosley}}, \bibinfo
  {author} {\bibfnamefont {E.}~\bibnamefont {Andersson}}, \bibinfo {author}
  {\bibfnamefont {I.}~\bibnamefont {Jex}}, \ and\ \bibinfo {author}
  {\bibfnamefont {C.}~\bibnamefont {Silberhorn}},\ }\href {\doibase 10.1103/PhysRevLett.104.050502} 
  {\bibfield {journal} {\bibinfo  {journal} {Phys. Rev. Lett.}\ }\textbf {\bibinfo
  {volume} {104}},\ \bibinfo {pages} {050502} (\bibinfo {year}
  {2010})}\BibitemShut {NoStop}%
\bibitem [{\citenamefont {Schreiber}\ \emph {et~al.}(2011)\citenamefont
  {Schreiber}, \citenamefont {Cassemiro}, \citenamefont {Poto{\v c}ek},
  \citenamefont {G{\'a}bris}, \citenamefont {Jex},\ and\ \citenamefont
  {Silberhorn}}]{sch-prl-106-180403}%
  \BibitemOpen
  \bibfield  {author} {\bibinfo {author} {\bibfnamefont {A.}~\bibnamefont
  {Schreiber}}, \bibinfo {author} {\bibfnamefont {K.~N.}\ \bibnamefont
  {Cassemiro}}, \bibinfo {author} {\bibfnamefont {V.}~\bibnamefont {Poto{\v
  c}ek}}, \bibinfo {author} {\bibfnamefont {A.}~\bibnamefont {G{\'a}bris}},
  \bibinfo {author} {\bibfnamefont {I.}~\bibnamefont {Jex}}, \ and\ \bibinfo
  {author} {\bibfnamefont {C.}~\bibnamefont {Silberhorn}},\ }\href {\doibase 10.1103/PhysRevLett.106.180403}
  {\bibfield  {journal} {\bibinfo  {journal} {Phys. Rev. Lett.}\ }\textbf
  {\bibinfo {volume} {106}},\ \bibinfo {pages} {180403} (\bibinfo {year}
  {2011})}\BibitemShut {NoStop}%
\bibitem [{\citenamefont {Perets}\ \emph {et~al.}(2008)\citenamefont {Perets},
  \citenamefont {Lahini}, \citenamefont {Pozzi}, \citenamefont {Sorel},
  \citenamefont {Morandotti},\ and\ \citenamefont
  {Silberberg}}]{pe-prl-100-170506}%
  \BibitemOpen
  \bibfield  {author} {\bibinfo {author} {\bibfnamefont {H.~B.}\ \bibnamefont
  {Perets}}, \bibinfo {author} {\bibfnamefont {Y.}~\bibnamefont {Lahini}},
  \bibinfo {author} {\bibfnamefont {F.}~\bibnamefont {Pozzi}}, \bibinfo
  {author} {\bibfnamefont {M.}~\bibnamefont {Sorel}}, \bibinfo {author}
  {\bibfnamefont {R.}~\bibnamefont {Morandotti}}, \ and\ \bibinfo {author}
  {\bibfnamefont {Y.}~\bibnamefont {Silberberg}},\ }\href {\doibase 10.1103/PhysRevLett.100.170506} 
  {\bibfield {journal} {\bibinfo  {journal} {Phys. Rev. Lett.}\ }\textbf {\bibinfo
  {volume} {100}},\ \bibinfo {pages} {170506} (\bibinfo {year}
  {2008})}\BibitemShut {NoStop}%
\bibitem [{\citenamefont {Knight}\ \emph {et~al.}(2003)\citenamefont {Knight},
  \citenamefont {Rold{\'a}n},\ and\ \citenamefont {Sipe}}]{kn-pra-68-020301}%
  \BibitemOpen
  \bibfield  {author} {\bibinfo {author} {\bibfnamefont {P.~L.}\ \bibnamefont
  {Knight}}, \bibinfo {author} {\bibfnamefont {E.}~\bibnamefont {Rold{\'a}n}},
  \ and\ \bibinfo {author} {\bibfnamefont {J.~E.}\ \bibnamefont {Sipe}},\
  }\href {\doibase 10.1103/PhysRevA.68.020301} {\bibfield  {journal} {\bibinfo  {journal} {Phys. Rev. A}\
  }\textbf {\bibinfo {volume} {68}},\ \bibinfo {pages} {020301(R)} (\bibinfo
  {year} {2003})}\BibitemShut {NoStop}%
\bibitem [{\citenamefont {Bromberg}\ \emph {et~al.}(2009)\citenamefont
  {Bromberg}, \citenamefont {Lahini}, \citenamefont {Morandotti},\ and\
  \citenamefont {Silberberg}}]{br-prl-102-253904}%
  \BibitemOpen
  \bibfield  {author} {\bibinfo {author} {\bibfnamefont {Y.}~\bibnamefont
  {Bromberg}}, \bibinfo {author} {\bibfnamefont {Y.}~\bibnamefont {Lahini}},
  \bibinfo {author} {\bibfnamefont {R.}~\bibnamefont {Morandotti}}, \ and\
  \bibinfo {author} {\bibfnamefont {Y.}~\bibnamefont {Silberberg}},\
  }\href {\doibase 10.1103/PhysRevLett.102.253904} {\bibfield  {journal} {\bibinfo  {journal} {Phys. Rev. Lett.}\
  }\textbf {\bibinfo {volume} {102}},\ \bibinfo {pages} {253904} (\bibinfo
  {year} {2009})}\BibitemShut {NoStop}%
\bibitem [{\citenamefont {Keil}\ \emph {et~al.}(2010)\citenamefont {Keil},
  \citenamefont {Szameit}, \citenamefont {Dreisow}, \citenamefont {Heinrich},
  \citenamefont {Nolte},\ and\ \citenamefont
  {T{\"u}nnermann}}]{ke-pra-81-023834}%
  \BibitemOpen
  \bibfield  {author} {\bibinfo {author} {\bibfnamefont {R.}~\bibnamefont
  {Keil}}, \bibinfo {author} {\bibfnamefont {A.}~\bibnamefont {Szameit}},
  \bibinfo {author} {\bibfnamefont {F.}~\bibnamefont {Dreisow}}, \bibinfo
  {author} {\bibfnamefont {M.}~\bibnamefont {Heinrich}}, \bibinfo {author}
  {\bibfnamefont {S.}~\bibnamefont {Nolte}}, \ and\ \bibinfo {author}
  {\bibfnamefont {A.}~\bibnamefont {T{\"u}nnermann}},\ }\href {\doibase 10.1103/PhysRevA.81.023834} 
  {\bibfield  {journal} {\bibinfo  {journal} {Phys. Rev. A}\ }\textbf {\bibinfo {volume} {81}},\
  \bibinfo {pages} {023834} (\bibinfo {year} {2010})}\BibitemShut {NoStop}%
\bibitem [{\citenamefont {Keil}\ \emph {et~al.}(2011)\citenamefont {Keil},
  \citenamefont {Dreisow}, \citenamefont {Heinrich}, \citenamefont
  {T{\"u}nnermann}, \citenamefont {Nolte},\ and\ \citenamefont
  {Szameit}}]{ke-pra-83-013808}%
  \BibitemOpen
  \bibfield  {author} {\bibinfo {author} {\bibfnamefont {R.}~\bibnamefont
  {Keil}}, \bibinfo {author} {\bibfnamefont {F.}~\bibnamefont {Dreisow}},
  \bibinfo {author} {\bibfnamefont {M.}~\bibnamefont {Heinrich}}, \bibinfo
  {author} {\bibfnamefont {A.}~\bibnamefont {T{\"u}nnermann}}, \bibinfo
  {author} {\bibfnamefont {S.}~\bibnamefont {Nolte}}, \ and\ \bibinfo {author}
  {\bibfnamefont {A.}~\bibnamefont {Szameit}},\ }\href {\doibase 10.1103/PhysRevA.83.013808} 
  {\bibfield {journal} {\bibinfo  {journal} {Phys. Rev. A}\ }\textbf {\bibinfo {volume}
  {83}},\ \bibinfo {pages} {013808} (\bibinfo {year} {2011})}\BibitemShut
  {NoStop}%
\bibitem [{\citenamefont {Schreiber}\ \emph {et~al.}(2012)\citenamefont
  {Schreiber}, \citenamefont {G{\'a}bris}, \citenamefont {Rohde}, \citenamefont
  {Laiho}, \citenamefont {{\v{S}}tefa{\v{n}}{\'a}k}, \citenamefont
  {Poto{\v{c}}ek}, \citenamefont {Hamilton}, \citenamefont {Jex},\ and\
  \citenamefont {Silberhorn}}]{sc-sci-336-55}%
  \BibitemOpen
  \bibfield  {author} {\bibinfo {author} {\bibfnamefont {A.}~\bibnamefont
  {Schreiber}}, \bibinfo {author} {\bibfnamefont {A.}~\bibnamefont
  {G{\'a}bris}}, \bibinfo {author} {\bibfnamefont {P.~P.}\ \bibnamefont {Rohde}},
  \bibinfo {author} {\bibfnamefont {K.}~\bibnamefont {Laiho}}, \bibinfo
  {author} {\bibfnamefont {M.}~\bibnamefont {{\v{S}}tefa{\v{n}}{\'a}k}},
  \bibinfo {author} {\bibfnamefont {V.}~\bibnamefont {Poto{\v{c}}ek}}, \bibinfo
  {author} {\bibfnamefont {C.}~\bibnamefont {Hamilton}}, \bibinfo {author}
  {\bibfnamefont {I.}~\bibnamefont {Jex}}, \ and\ \bibinfo {author}
  {\bibfnamefont {C.}~\bibnamefont {Silberhorn}},\ }\href {\doibase 10.1126/science.1218448} 
  {\bibfield {journal} {\bibinfo  {journal} {Science}\ }\textbf {\bibinfo {volume}
  {336}},\ \bibinfo {pages} {55} (\bibinfo {year} {2012})}\BibitemShut
  {NoStop}%
\bibitem [{\citenamefont {Rohde}\ \emph {et~al.}(2012)\citenamefont {Rohde},
  \citenamefont {Schreiber}, \citenamefont {Stefanak}, \citenamefont {Jex},
  \citenamefont {Gilchrist},\ and\ \citenamefont
  {Silberhorn}}]{ro-arxiv-1205.1850}%
  \BibitemOpen
  \bibfield  {author} {\bibinfo {author} {\bibfnamefont {P.~P.}\ \bibnamefont
  {Rohde}}, \bibinfo {author} {\bibfnamefont {A.}~\bibnamefont {Schreiber}},
  \bibinfo {author} {\bibfnamefont {M.}~\bibnamefont {Stefanak}}, \bibinfo
  {author} {\bibfnamefont {I.}~\bibnamefont {Jex}}, \bibinfo {author}
  {\bibfnamefont {A.}~\bibnamefont {Gilchrist}}, \ and\ \bibinfo {author}
  {\bibfnamefont {C.}~\bibnamefont {Silberhorn}},\ }\href {http://arxiv.org/abs/1205.1850} 
  {\bibfield {journal} {\bibinfo  {journal} {arXiv:1205.1850}\ } (\bibinfo {year}
  {2012})}\BibitemShut {NoStop}%
\bibitem [{\citenamefont
  {Longhi}(2011)}]{Longhi:BlochOscillationsCorrelatedParticlesSim2D}%
  \BibitemOpen
  \bibfield  {author} {\bibinfo {author} {\bibfnamefont {S.}~\bibnamefont
  {Longhi}},\ }\href {\doibase 10.1364/OL.36.003248} {\bibfield  {journal} {\bibinfo  {journal} {Opt.
  Lett.}\ }\textbf {\bibinfo {volume} {36}},\ \bibinfo {pages} {3248} (\bibinfo
  {year} {2011})}\BibitemShut {NoStop}%
\bibitem [{\citenamefont {Gr\"{a}fe}\ \emph {et~al.}(2012)\citenamefont
  {Gr\"{a}fe}, \citenamefont {Solntsev}, \citenamefont {Keil}, \citenamefont
  {Sukhorukov}, \citenamefont {Heinrich}, \citenamefont {T\"{u}nnermann},
  \citenamefont {Nolte}, \citenamefont {Szameit},\ and\ \citenamefont
  {Kivshar}}]{Graefe:2DSimulatorBiphotonWalk}%
  \BibitemOpen
  \bibfield  {author} {\bibinfo {author} {\bibfnamefont {M.}~\bibnamefont
  {Gr\"{a}fe}}, \bibinfo {author} {\bibfnamefont {A.~S.}\ \bibnamefont
  {Solntsev}}, \bibinfo {author} {\bibfnamefont {R.}~\bibnamefont {Keil}},
  \bibinfo {author} {\bibfnamefont {A.~A.}\ \bibnamefont {Sukhorukov}},
  \bibinfo {author} {\bibfnamefont {M.}~\bibnamefont {Heinrich}}, \bibinfo
  {author} {\bibfnamefont {A.}~\bibnamefont {T\"{u}nnermann}}, \bibinfo
  {author} {\bibfnamefont {S.}~\bibnamefont {Nolte}}, \bibinfo {author}
  {\bibfnamefont {A.}~\bibnamefont {Szameit}}, \ and\ \bibinfo {author}
  {\bibfnamefont {Y.~S.}\ \bibnamefont {Kivshar}},\ }\href {\doibase 10.1038/srep00562} 
  {\bibfield {journal} {\bibinfo  {journal} {Sci. Rep.}\ }\textbf {\bibinfo {volume}
  {2}},\ \bibinfo {pages} {562} (\bibinfo {year} {2012})}\BibitemShut {NoStop}%
\bibitem [{\citenamefont {Mohseni}\ \emph {et~al.}(2008)\citenamefont
  {Mohseni}, \citenamefont {Rebentrost}, \citenamefont {Lloyd},\ and\
  \citenamefont {Guzik}}]{mo-jchemphys-129-174106}%
  \BibitemOpen
  \bibfield  {author} {\bibinfo {author} {\bibfnamefont {M.}~\bibnamefont
  {Mohseni}}, \bibinfo {author} {\bibfnamefont {P.}~\bibnamefont {Rebentrost}},
  \bibinfo {author} {\bibfnamefont {S.}~\bibnamefont {Lloyd}}, \ and\ \bibinfo
  {author} {\bibfnamefont {A.}\ \bibnamefont {Aspuru-Guzik}},\ }\href {\doibase 10.1063/1.3002335} 
  {\bibfield  {journal} {\bibinfo  {journal} {J. Chem. Phys.}\ }\textbf {\bibinfo {volume} {129}},\ \bibinfo {pages}
  {174106+} (\bibinfo {year} {2008})}\BibitemShut {NoStop}%
\bibitem [{\citenamefont {Harrison}\ \emph {et~al.}(2011)\citenamefont
  {Harrison}, \citenamefont {Keating},\ and\ \citenamefont
  {Robbins}}]{ha-prsa-467-212}%
  \BibitemOpen
  \bibfield  {author} {\bibinfo {author} {\bibfnamefont {J.~M.}\ \bibnamefont
  {Harrison}}, \bibinfo {author} {\bibfnamefont {J.~P.}\ \bibnamefont
  {Keating}}, \ and\ \bibinfo {author} {\bibfnamefont {J.~M.}\ \bibnamefont
  {Robbins}},\ }\href {\doibase 10.1098/rspa.2010.0254} {\bibfield  {journal}
  {\bibinfo  {journal} {Proc. R. Soc. A}\ }\textbf {\bibinfo {volume} {467}},\
  \bibinfo {pages} {212} (\bibinfo {year} {2011})}\BibitemShut {NoStop}%
\bibitem [{\citenamefont {Childs}\ \emph {et~al.}(2003)\citenamefont {Childs},
  \citenamefont {Cleve}, \citenamefont {Deotto}, \citenamefont {Farhi},
  \citenamefont {Gutmann},\ and\ \citenamefont {Spielman}}]{ch-stoc03proc-59}%
  \BibitemOpen
  \bibfield  {author} {\bibinfo {author} {\bibfnamefont {A.~M.}\ \bibnamefont
  {Childs}}, \bibinfo {author} {\bibfnamefont {R.}~\bibnamefont {Cleve}},
  \bibinfo {author} {\bibfnamefont {E.}~\bibnamefont {Deotto}}, \bibinfo
  {author} {\bibfnamefont {E.}~\bibnamefont {Farhi}}, \bibinfo {author}
  {\bibfnamefont {S.}~\bibnamefont {Gutmann}}, \ and\ \bibinfo {author}
  {\bibfnamefont {D.~A.}\ \bibnamefont {Spielman}},\ }in\ \href@noop {} {\emph {\bibinfo {booktitle} {Proceedings of the
  thirty-fifth annual ACM symposium on Theory of computing}}},\ \bibinfo
  {series and number} {STOC '03}\ (\bibinfo  {publisher} {ACM},\ \bibinfo
  {address} {New York, NY, USA},\ \bibinfo {year} {2003})\ pp.\ \bibinfo
  {pages} {59--68}\BibitemShut {NoStop}%
\bibitem [{\citenamefont {Marshall}\ \emph {et~al.}(2009)\citenamefont
  {Marshall}, \citenamefont {Politi}, \citenamefont {Matthews}, \citenamefont
  {Dekker}, \citenamefont {Ams}, \citenamefont {Withford},\ and\ \citenamefont
  {O'Brien}}]{ma-oe-17-12546}%
  \BibitemOpen
  \bibfield  {author} {\bibinfo {author} {\bibfnamefont {G.~D.}\ \bibnamefont
  {Marshall}}, \bibinfo {author} {\bibfnamefont {A.}~\bibnamefont {Politi}},
  \bibinfo {author} {\bibfnamefont {J.~C.~F.}\ \bibnamefont {Matthews}},
  \bibinfo {author} {\bibfnamefont {P.}~\bibnamefont {Dekker}}, \bibinfo
  {author} {\bibfnamefont {M.}~\bibnamefont {Ams}}, \bibinfo {author}
  {\bibfnamefont {M.~J.}\ \bibnamefont {Withford}}, \ and\ \bibinfo {author}
  {\bibfnamefont {J.~L.}\ \bibnamefont {O'Brien}},\ }\href {\doibase 10.1364/OE.17.012546} 
  {\bibfield {journal} {\bibinfo  {journal} {Opt. Express}\ }\textbf {\bibinfo {volume}
  {17}},\ \bibinfo {pages} {12546} (\bibinfo {year} {2009})}\BibitemShut
  {NoStop}%
\bibitem [{\citenamefont {Sansoni}\ \emph {et~al.}(2012)\citenamefont
  {Sansoni}, \citenamefont {Sciarrino}, \citenamefont {Vallone}, \citenamefont
  {Mataloni}, \citenamefont {Crespi}, \citenamefont {Ramponi},\ and\
  \citenamefont {Osellame}}]{sa-prl-108-010502}%
  \BibitemOpen
  \bibfield  {author} {\bibinfo {author} {\bibfnamefont {L.}~\bibnamefont
  {Sansoni}}, \bibinfo {author} {\bibfnamefont {F.}~\bibnamefont {Sciarrino}},
  \bibinfo {author} {\bibfnamefont {G.}~\bibnamefont {Vallone}}, \bibinfo
  {author} {\bibfnamefont {P.}~\bibnamefont {Mataloni}}, \bibinfo {author}
  {\bibfnamefont {A.}~\bibnamefont {Crespi}}, \bibinfo {author} {\bibfnamefont
  {R.}~\bibnamefont {Ramponi}}, \ and\ \bibinfo {author} {\bibfnamefont
  {R.}~\bibnamefont {Osellame}},\ }\href {\doibase 10.1103/PhysRevLett.108.010502} {\bibfield  {journal}
  {\bibinfo  {journal} {Phys. Rev. Lett.}\ }\textbf {\bibinfo {volume} {108}},\
  \bibinfo {pages} {010502} (\bibinfo {year} {2012})}\BibitemShut {NoStop}%
\bibitem [{\citenamefont {Owens}\ \emph {et~al.}(2011)\citenamefont {Owens},
  \citenamefont {Broome}, \citenamefont {Biggerstaff}, \citenamefont {Goggin},
  \citenamefont {Fedrizzi}, \citenamefont {Linjordet}, \citenamefont {Ams},
  \citenamefont {Marshall}, \citenamefont {Twamley}, \citenamefont {Withford},\
  and\ \citenamefont {White}}]{ow-njp-13-075003}%
  \BibitemOpen
  \bibfield  {author} {\bibinfo {author} {\bibfnamefont {J.~O.}\ \bibnamefont
  {Owens}}, \bibinfo {author} {\bibfnamefont {M.~A.}\ \bibnamefont {Broome}},
  \bibinfo {author} {\bibfnamefont {D.~N.}\ \bibnamefont {Biggerstaff}},
  \bibinfo {author} {\bibfnamefont {M.~E.}\ \bibnamefont {Goggin}}, \bibinfo
  {author} {\bibfnamefont {A.}~\bibnamefont {Fedrizzi}}, \bibinfo {author}
  {\bibfnamefont {T.}~\bibnamefont {Linjordet}}, \bibinfo {author}
  {\bibfnamefont {M.}~\bibnamefont {Ams}}, \bibinfo {author} {\bibfnamefont
  {G.~D.}\ \bibnamefont {Marshall}}, \bibinfo {author} {\bibfnamefont
  {J.}~\bibnamefont {Twamley}}, \bibinfo {author} {\bibfnamefont {M.~J.}\
  \bibnamefont {Withford}}, \ and\ \bibinfo {author} {\bibfnamefont {A.~G.}\
  \bibnamefont {White}},\ }\href {\doibase 10.1088/1367-2630/13/7/075003} {\bibfield  {journal} {\bibinfo
  {journal} {New J. Phys.}\ }\textbf {\bibinfo {volume} {13}},\ \bibinfo
  {pages} {075003} (\bibinfo {year} {2011})}\BibitemShut {NoStop}%
\bibitem [{\citenamefont {Szameit}\ and\ \citenamefont
  {Nolte}(2010)}]{sz-jphysb-43-163001}%
  \BibitemOpen
  \bibfield  {author} {\bibinfo {author} {\bibfnamefont {A.}~\bibnamefont
  {Szameit}}\ and\ \bibinfo {author} {\bibfnamefont {S.}~\bibnamefont
  {Nolte}},\ }\href {\doibase 10.1088/0953-4075/43/16/163001} {\bibfield
  {journal} {\bibinfo  {journal} {J. Phys. B: At. Mol. Opt. Phys.}\ }\textbf {\bibinfo {volume} {43}},\ \bibinfo {pages}
  {163001+} (\bibinfo {year} {2010})}\BibitemShut {NoStop}%
\bibitem [{\citenamefont {Mattle}\ \emph {et~al.}(1995)\citenamefont {Mattle},
  \citenamefont {Michler}, \citenamefont {Weinfurter}, \citenamefont
  {Zeilinger},\ and\ \citenamefont {Zukowski}}]{ma-apb-60-s111}%
  \BibitemOpen
  \bibfield  {author} {\bibinfo {author} {\bibfnamefont {K.}~\bibnamefont
  {Mattle}}, \bibinfo {author} {\bibfnamefont {M.}~\bibnamefont {Michler}},
  \bibinfo {author} {\bibfnamefont {H.}~\bibnamefont {Weinfurter}}, \bibinfo
  {author} {\bibfnamefont {A.}~\bibnamefont {Zeilinger}}, \ and\ \bibinfo
  {author} {\bibfnamefont {M.}~\bibnamefont {Zukowski}},\ }\href@noop {}
  {\bibfield  {journal} {\bibinfo  {journal} {Appl. Phys. B}\ }\textbf
  {\bibinfo {volume} {60}},\ \bibinfo {pages} {S111} (\bibinfo {year}
  {1995})}\BibitemShut {NoStop}%
\bibitem [{\citenamefont {Peruzzo}\ \emph {et~al.}(2011)\citenamefont
  {Peruzzo}, \citenamefont {Laing}, \citenamefont {Politi}, \citenamefont
  {Rudolph},\ and\ \citenamefont {O'Brien}}]{pe-ncomms-2-224}%
  \BibitemOpen
  \bibfield  {author} {\bibinfo {author} {\bibfnamefont {A.}~\bibnamefont
  {Peruzzo}}, \bibinfo {author} {\bibfnamefont {A.}~\bibnamefont {Laing}},
  \bibinfo {author} {\bibfnamefont {A.}~\bibnamefont {Politi}}, \bibinfo
  {author} {\bibfnamefont {T.}~\bibnamefont {Rudolph}}, \ and\ \bibinfo
  {author} {\bibfnamefont {J.~L.}\ \bibnamefont {O'Brien}},\ }\href {\doibase 10.1038/ncomms1228} 
  {\bibfield  {journal} {\bibinfo  {journal} {Nat. Commun.}\ }\textbf {\bibinfo {volume} {2}},\ \bibinfo {pages} {224+}
  (\bibinfo {year} {2011})}\BibitemShut {NoStop}%
\bibitem [{\citenamefont {Matthews}\ \emph {et~al.}(2013)\citenamefont
  {Matthews}, \citenamefont {Poulios}, \citenamefont {Meinecke}, \citenamefont
  {Politi}, \citenamefont {Peruzzo}, \citenamefont {Ismail}, \citenamefont
  {W\"{o}rhoff}, \citenamefont {Thompson},\ and\ \citenamefont
  {O'Brien}}]{ma-scirep-3-1539}%
  \BibitemOpen
  \bibfield  {author} {\bibinfo {author} {\bibfnamefont {J.~C.~F.}\
  \bibnamefont {Matthews}}, \bibinfo {author} {\bibfnamefont {K.}~\bibnamefont
  {Poulios}}, \bibinfo {author} {\bibfnamefont {J.~D.~A.}\ \bibnamefont
  {Meinecke}}, \bibinfo {author} {\bibfnamefont {A.}~\bibnamefont {Politi}},
  \bibinfo {author} {\bibfnamefont {A.}~\bibnamefont {Peruzzo}}, \bibinfo
  {author} {\bibfnamefont {N.}~\bibnamefont {Ismail}}, \bibinfo {author}
  {\bibfnamefont {K.}~\bibnamefont {W\"{o}rhoff}}, \bibinfo {author}
  {\bibfnamefont {M.~G.}\ \bibnamefont {Thompson}}, \ and\ \bibinfo {author}
  {\bibfnamefont {J.~L.}\ \bibnamefont {O'Brien}},\ }\href {\doibase 10.1038/srep01539} 
  {\bibfield  {journal} {\bibinfo  {journal} {Sci. Rep.}\ }\textbf {\bibinfo {volume} {3}},\ \bibinfo {pages} {1539} (\bibinfo {year}
  {2013})}\BibitemShut {NoStop}%
\bibitem [{\citenamefont {Meinecke}\ \emph {et~al.}(2013)\citenamefont
  {Meinecke}, \citenamefont {Poulios}, \citenamefont {Politi}, \citenamefont
  {Matthews}, \citenamefont {Peruzzo}, \citenamefont {Ismail}, \citenamefont
  {W\"{o}rhoff}, \citenamefont {O'Brien},\ and\ \citenamefont
  {Thompson}}]{me-pra-88-012308}%
  \BibitemOpen
  \bibfield  {author} {\bibinfo {author} {\bibfnamefont {J.~D.~A.}\
  \bibnamefont {Meinecke}}, \bibinfo {author} {\bibfnamefont {K.}~\bibnamefont
  {Poulios}}, \bibinfo {author} {\bibfnamefont {A.}~\bibnamefont {Politi}},
  \bibinfo {author} {\bibfnamefont {J.~C.~F.}\ \bibnamefont {Matthews}},
  \bibinfo {author} {\bibfnamefont {A.}~\bibnamefont {Peruzzo}}, \bibinfo
  {author} {\bibfnamefont {N.}~\bibnamefont {Ismail}}, \bibinfo {author}
  {\bibfnamefont {K.}~\bibnamefont {W\"{o}rhoff}}, \bibinfo {author}
  {\bibfnamefont {J.~L.}\ \bibnamefont {O'Brien}}, \ and\ \bibinfo {author}
  {\bibfnamefont {M.~G.}\ \bibnamefont {Thompson}},\ }\href {\doibase
  10.1103/physreva.88.012308} {\bibfield  {journal} {\bibinfo  {journal}
  {Phys. Rev. A}\ }\textbf {\bibinfo {volume} {88}},\ \bibinfo {pages}
  {012308+} (\bibinfo {year} {2013})}\BibitemShut {NoStop}%
\bibitem [{\citenamefont {Knill}\ \emph {et~al.}(2001)\citenamefont {Knill},
  \citenamefont {Laflamme},\ and\ \citenamefont {Milburn}}]{kn-nature-409-46}%
  \BibitemOpen
  \bibfield  {author} {\bibinfo {author} {\bibfnamefont {E.}~\bibnamefont
  {Knill}}, \bibinfo {author} {\bibfnamefont {R.}~\bibnamefont {Laflamme}}, \
  and\ \bibinfo {author} {\bibfnamefont {G.~J.}\ \bibnamefont {Milburn}},\
  }\href {\doibase 10.1038/35051009} {\bibfield  {journal} {\bibinfo  {journal}
  {Nature}\ }\textbf {\bibinfo {volume} {409}},\ \bibinfo {pages} {46}
  (\bibinfo {year} {2001})}\BibitemShut {NoStop}%
\bibitem [{\citenamefont {Corrielli}\ \emph {et~al.}(2013)\citenamefont
  {Corrielli}, \citenamefont {Crespi}, \citenamefont {Della~Valle},
  \citenamefont {Longhi},\ and\ \citenamefont {Osellame}}]{co-ncomm-4-1555}%
  \BibitemOpen
  \bibfield  {author} {\bibinfo {author} {\bibfnamefont {G.}~\bibnamefont
  {Corrielli}}, \bibinfo {author} {\bibfnamefont {A.}~\bibnamefont {Crespi}},
  \bibinfo {author} {\bibfnamefont {G.}~\bibnamefont {Della~Valle}}, \bibinfo
  {author} {\bibfnamefont {S.}~\bibnamefont {Longhi}}, \ and\ \bibinfo {author}
  {\bibfnamefont {R.}~\bibnamefont {Osellame}},\ }\href {\doibase 10.1038/ncomms2578} 
  {\bibfield  {journal} {\bibinfo  {journal} {Nat. Commun.}\ 
  }\textbf {\bibinfo {volume} {4}},\ \bibinfo {pages} {1555+} (\bibinfo {year}
  {2013})}\BibitemShut {NoStop}%
\bibitem [{\citenamefont {Childs}\ \emph {et~al.}(2013)\citenamefont {Childs},
  \citenamefont {Gosset},\ and\ \citenamefont {Webb}}]{ch-sci-339-791}%
  \BibitemOpen
  \bibfield  {author} {\bibinfo {author} {\bibfnamefont {A.~M.}\ \bibnamefont
  {Childs}}, \bibinfo {author} {\bibfnamefont {D.}~\bibnamefont {Gosset}}, \
  and\ \bibinfo {author} {\bibfnamefont {Z.}~\bibnamefont {Webb}},\ }\href {\doibase 10.1126/science.1229957 } 
  {\bibfield  {journal} {\bibinfo  {journal}
  {Science}\ }\textbf {\bibinfo {volume} {339}},\ \bibinfo {pages} {791}
  (\bibinfo {year} {2013})}\BibitemShut {NoStop}%
\end{thebibliography}
\end{document}